\documentclass{PoS}

\usepackage{multirow, graphicx, epsfig, color}
   \usepackage{amsmath}
   \usepackage{amssymb}
   \usepackage{epstopdf}
   \usepackage{pdfsync}
   \usepackage{gensymb}

\DeclareGraphicsExtensions{.pdf}

\def\teq#1{$\, #1\,$}                         
\def\edithere#1{\textcolor{red}{#1}}  

\font\fiverm=cmr5             \font\sevenrm=cmr7

          \font\sixrm=cmr6       

\def\pmb#1{\setbox0=\hbox{#1}%
  \kern-0.0125em\copy0\kern-\wd0
  \kern0.025em\copy0\kern-\wd0
  \kern-0.0125em\raise0.0433em\box0 }

\def\erg{\varepsilon}

\def\fsc{\alpha_{\hbox{\sevenrm f}}}  
\def\sigt{\sigma_{\hbox{\fiverm T}}}
\def\taut{\tau_{\hbox{\fiverm T}}}

\def\rns{R_{\hbox{\sixrm NS}}}

\def\rmax{r_{\hbox{\sevenrm max}}} 
\def\eesc{\erg_{\hbox{\sevenrm esc}}} 
\def\thetakB{\theta_{\hbox{\sevenrm kB}}} 
\def\muvec{\pmb{$\mu$}} 
\def\Omegavec{\pmb{$\Omega$}} 
\def\eescsp{\varepsilon^{sp}_{\rm esc}}
\def\efmax{\varepsilon_f^{\rm max}}
\def\nGJ{n_{\hbox{\sixrm GJ}}} 

\def\Chandra{{\sl Chandra}}
\def\RXTE{{\sl RXTE}}
\def\INTEGRAL{{\sl INTEGRAL}}
\def\COMPTEL{{\sl COMPTEL}}
\def\NuSTAR{{\sl NuSTAR}}
\def\Suzaku{{\sl Suzaku}}

\def\acknowledgements#1{\noindent{\bf Acknowledgements}: #1}

\newcommand{\vol}[2]{$\,$\bf #1\rm , #2}                 

\newcommand{\figureoutpdf}[5]{
   \centerline{\hspace{#3in} \includegraphics[width=#2truein]{#1}}
   \vspace{#4truein} \caption{#5}  \vspace{-0.35truein}}

\title{The Mysterious Magnetospheres of Magnetars}

\ShortTitle{The Mysterious Magnetospheres of Magnetars}

\author{\speaker{Matthew G. Baring}\\ 
        Department of Physics and Astronomy - MS 108, Rice University,
        6100 Main Street, Houston, Texas 77251-1892, USA\\
        E-mail: \email{baring@rice.edu}}

\author{Zorawar Wadiasingh\\
	Gravitational Astrophysics Laboratory, Code 663, 
	NASA's Goddard Space Flight Center, Greenbelt, Maryland, 20771, USA
	\thanks{Universities Space Research Association, Columbia, MD 21046, USA}\\
        E-mail: \email{zwadiasingh@gmail.com}}

\author{Peter L. Gonthier\\
	Hope College, Department of Physics, 27 Graves Place, Holland, MI 49423, USA\\
        E-mail: \email{gonthier@hope.edu}}

\author{Alice K. Harding\\
	Theoretical Division, Los Alamos National Laboratory, Los Alamos, NM 87545, USA\\
        E-mail: \email{ahardingx@yahoo.com}}

\author{Kun Hu\\
        Department of Physics and Astronomy - MS 108, Rice University,
        6100 Main Street, Houston, Texas 77251-1892, USA\\
        E-mail: \email{kh38@rice.edu}}

\abstract{
Magnetars are the most luminous compact objects in the stellar mass
range observed in the Milky Way, with giant flares of hard X-ray power \teq{\gtrsim 10^{45}}erg/sec
being detected from three soft gamma repeaters in the Galactic neighborhood.  
Periodicity seen in magnetar persistent 
emission, and a distinctive "spin-down" lengthening of this
period, have driven the paradigm that strongly-magnetized neutron stars
constitute these fascinating sources. The steady X-ray emission includes
both thermal atmospheric components, and magnetospheric contributions
that are manifested as hard X-ray ``tails.'' This paper addresses
observational and theoretical elements pertinent to the steady
hard X-ray emission of magnetars, focusing on dissipative processes in
their magnetospheres, and elements of Comptonization and polarization.
It also discusses the action and possible signatures of the exotic and fundamental QED
mechanisms of photon splitting and magnetic pair creation, and the
quest for their observational vindication.
}

\FullConference{7th Annual Conference on High Energy Astrophysics in Southern Africa\\
		28 - 30 August 2019\\
		Swakopmund, Namibia\\
		\edithere{In PoS on-line proceedings 371 [036]} } 

\begin{document}

\section{Introduction}

Magnetars have fascinated high-energy astrophysicists for four decades, propelled by
the first observation of a giant flare on 5th March 1979 
from the soft gamma repeater SGR 0525-66 \cite{Mazets_1981_ApSS}.  
They are highly-magnetized (\teq{B \gtrsim 10^{14}}Gauss) neutron stars that
have historically been divided into two observational groups: Soft-Gamma
Repeaters (SGRs) and Anomalous X-ray Pulsars (AXPs).  Their extreme fields 
are inferred directly from their timing properties presuming that their 
rapid rotational spin down is due to magnetic dipole torques 
(e.g. see \cite{Kouveliotou_1998_Nat}).  Such a class of neutron stars with 
superstrong fields was postulated as a model for SGRs by \cite{DT92}, 
and for AXPs by \cite{TD96}.  For reviews of magnetar science, see 
\cite{Mereghetti08,TZW15,Kaspi_2017_ARAA}.   Since their magnetic fields exceed
the quantum critical value of \teq{B_{\rm cr}=m_e^2c^3/(e\hbar)\approx 4.41\times
10^{13}}Gauss (the Schwinger value where the cyclotron energy \teq{\hbar (eB/m_ec)}
of the electron equals its rest mass energy \teq{m_ec^2}) the treatment of exotic processes in QED
is mandated.  This physics regime is not presently accessible by terrestrial 
experimental facilities, rendering cosmic magnetars an important
testbed for physics theory.

SGRs, over a dozen in number, are transients that exhibit repeated soft gamma-ray bursts
of subsecond duration in the 
\teq{10^{38}}erg/sec \teq{ < L < 10^{42}} erg/sec range, though three of 
them have exhibited giant super-second flares of energies exceeding
\teq{10^{45}}ergs (e.g. see \cite{Hurley99} for SGR 1900+14, and \cite{Palmer_2005_Nat} 
for SGR 1806-20), flares that could possibly be accompanied by gravitational wave signals 
detectable by aLIGO.  They also exhibit 
quiescent emission with periods \teq{P} in the range 2--12 sec. 
(e.g. see \cite{Kouveliotou_1998_Nat}, for SGR 1806-20).
The AXPs are a group of around a dozen pulsating, steady, bright X-ray sources with similar periods. 
Their quiescent signals are mostly thermal with steep power-law tails (e.g. \cite{Perna_2001_ApJ,Vigano13}),
and flat, hard X-ray tails (e.g., \cite{Kuiper04,Kuiper06,Hartog08a,Hartog08b})
that are the subject of this paper. 
AXPs possess persistent luminosities \teq{L_X \sim 10^{35}\,\rm erg\; s^{-1}}; 
as with the SGRs, these \teq{L_X} values far exceed their rotational power, so they
are possibly fueled by their internal magnetic energy.  Observations of
outburst activity in AXP 1E 2259+586 \cite{Gavriil_2004_ApJ},
in AXP 1E1841-045 \cite{Lin_2011_ApJ}, and in others suggest that anomalous X-ray
pulsars are indeed very similar to SGRs.  This ``unification paradigm''
has garnered widespread support within the magnetar community over the last decade.
The observational status quo of magnetars is summarized in the McGill Magnetar Catalog 
\cite{OK14}.\footnote{An on-line version is at 
{\it http://www.physics.mcgill.ca/\~{}pulsar/magnetar/main.html};
for a compendium of burst observational papers, see also {\it https://staff.fnwi.uva.nl/a.l.watts/magnetar/mb.html}.}

\begin{figure*}[ht]
\abovecaptionskip=0pt
\centering
\centerline{\includegraphics[width=5.75cm]{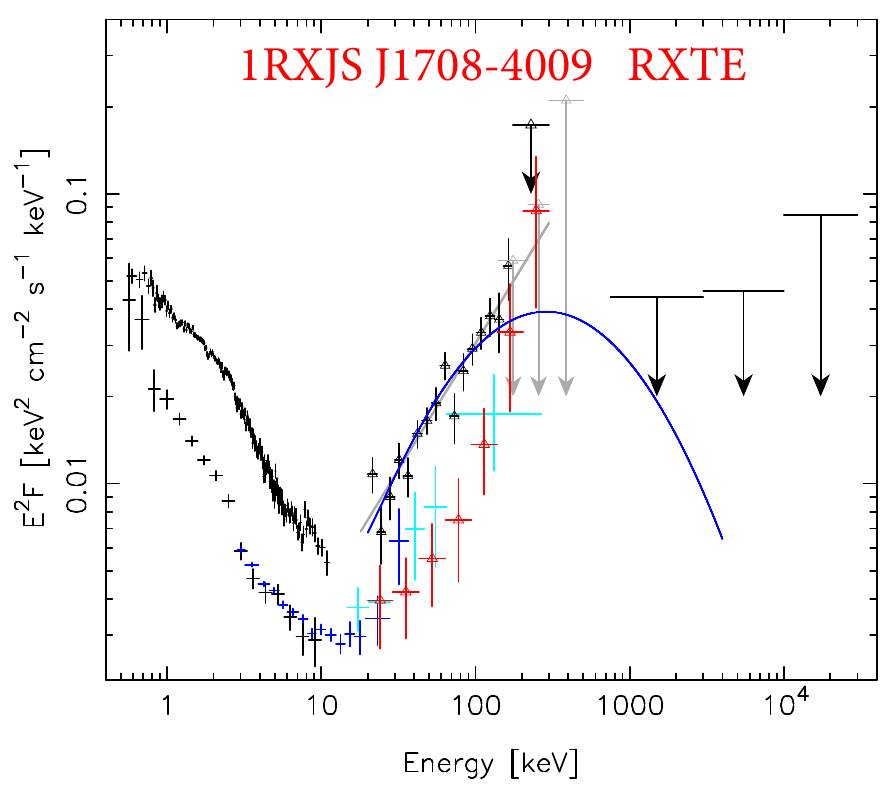}
	\hskip -0.35truecm \includegraphics[width=4.95cm]{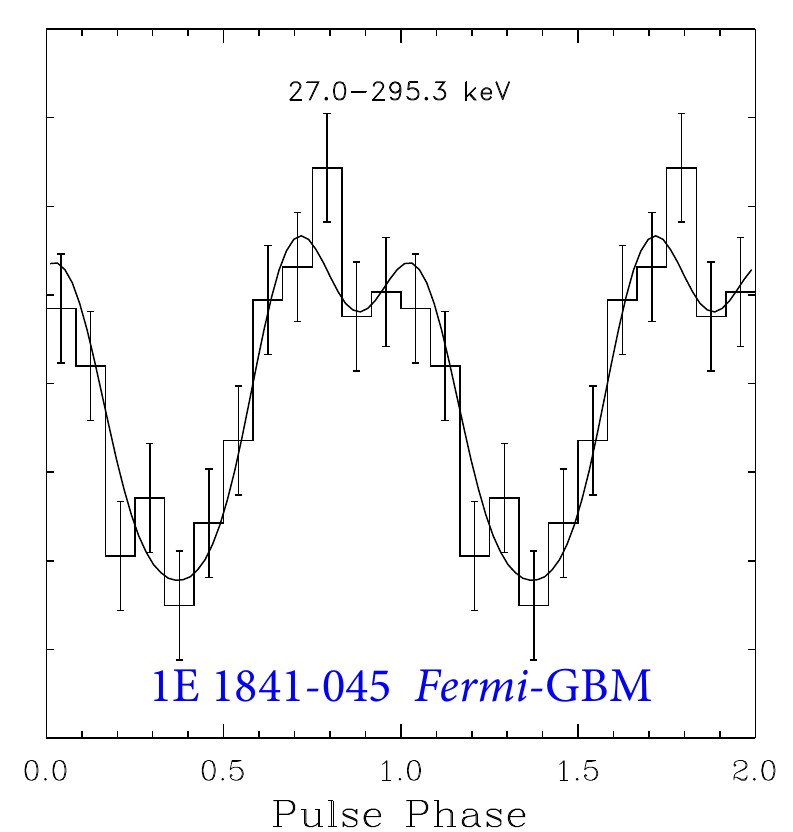}
	\hskip -0.3truecm \includegraphics[width=5.15cm]{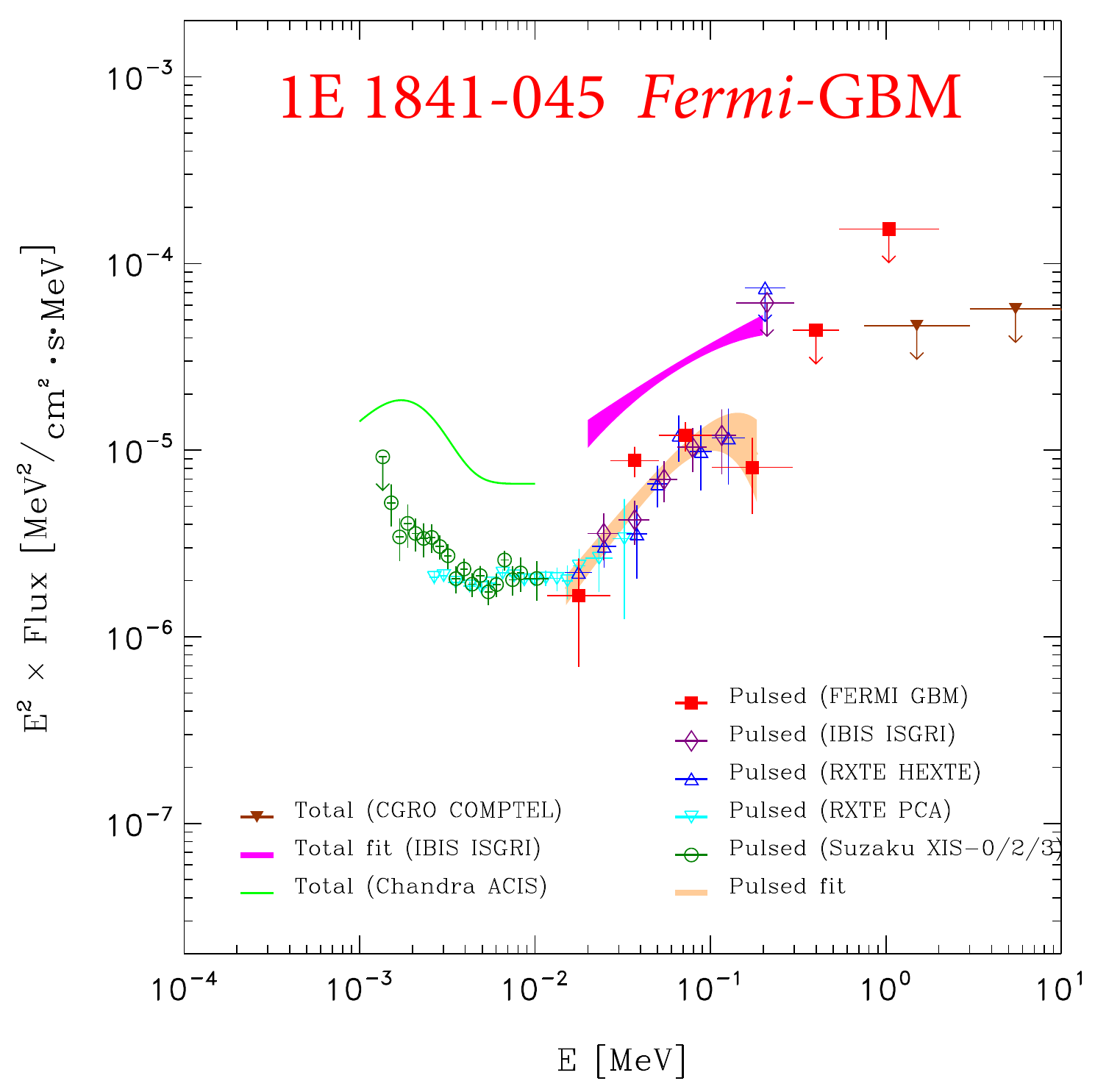}}
\vspace{0pt}
\caption{{\it Left panel}: The \teq{\nu F_{\nu}} X-ray spectrum of the AXP 1RXJS J1708-4009, with 
XMM data below 10 keV, and \RXTE-PCA/HEXTE data above 10 keV (red=pulsed) 
defining the hard X-ray tail.  The non-contemporaneous \COMPTEL\ upper 
limits above 1 MeV are also shown, as is an empirical spectral fit (blue)
-- see Fig.~4 of \cite{Hartog08b}.
{\it Middle panel}: Pulse profile of AXP 1E 1841-045 using {\it Fermi}-GBM data 
-- Fig.~18 of \cite{terBeek12}.  The smooth line shows the template profile observed 
by \INTEGRAL\ IBIS-ISGRI, 10-Mar-2003 to 30-Sep-2009, at 50-150 keV.  
{\it Right panel}: The spectrum of quasi-thermal (\teq{\lesssim 10}keV; surface) and tail 
(\teq{\gtrsim 10}keV; magnetosphere) quiescent emission from 1E 1841-045,
with pulsed emission represented by points: red is {\it Fermi}-GBM (maybe with a break 
at \teq{\sim 150} keV), 
and other points constitute \Chandra, \RXTE, \Suzaku\ and \INTEGRAL\ data.
\COMPTEL\  upper limits above 1 MeV are also shown.  See Fig.~25 in \cite{terBeek12}.}
 \label{fig:hard_tail_spec_pulse}      
\end{figure*}

The discovery by \INTEGRAL\ and \RXTE\ of hard, non-thermal pulsed spectral
tails in AXPs  \cite{Kuiper04,Kuiper06,Hartog08a,Hartog08b} added to 
the magnetar mystique by signalling the existence of a sustained magnetospheric 
component to their radiative resum\'{e}.
These luminous tails are extremely hard, typically extending up to 150 - 200 keV,
but with a turnover below around 500 keV implied by constraining {\it pre-2000} \COMPTEL\
upper limits (see Fig.~\ref{fig:hard_tail_spec_pulse}).  
Similar persistent emission tails are seen in SGRs (e.g. \cite{Goetz06} for SGR 1900+14).  
{\it Fermi}-GBM has also observed these tails \cite{terBeek12}, providing 
the sensitivity to better measure the flux above 100 keV (L. Kuiper, private communication).
The pulse profile and spectrum 
for AXP 1E 1841-045 are exhibited in Fig.~\ref{fig:hard_tail_spec_pulse}, with 
the latter suggesting possible evidence for a turnover at around 150 keV.
It is notable that this tail emission component, now seen in 9 magnetars, is not detected by the 
{\it Fermi}-LAT \cite{Abdo10_Fermi,Lietal2017}.  It is this hard X-ray component 
that is the subject of an ongoing investigation by our team, some details of 
which we present here as we progress along the labyrinthine path for demystifying 
the magnetospheres of magnetars.

\section{Hard X-ray Tail Modeling Essentials} 

\noindent{\bf Opacity}:
The most efficient means for producing the
hard tails in the 10--200 keV range is via resonant inverse Compton scattering (RICS) by 
energetic electrons.  This is the leading scenario for the production of this 
signal, where relativistic electrons energized in the inner magnetosphere 
upscatter surface thermal X rays of \teq{kT_s \sim 0.5}keV (\cite{BH07,FT07,Nobili08},
and later papers).  In this picture, Thomson optically thin conditions exist. 
To discern this, let \teq{{\cal E}_e\gtrsim L_{\gamma}/(4\pi R^2c)}
be the representative kinetic energy density in radiating electrons/pairs of mean 
Lorentz factor \teq{\langle\gamma_e\rangle \sim 10-100}.  Since the electron
number density is \teq{n_e \sim  {\cal E}_e/( \langle\gamma_e\rangle m_ec^2)}, 
one quickly arrives at the non-magnetic Thomson optical depth
\teq{\taut =n_e\sigt R\gtrsim L_{\gamma}\, \sigt/(4\pi Rm_ec^3\, \langle\gamma_e\rangle)}.  
For \teq{R\sim 10^6}cm and the observed persistent hard X-ray luminosities 
\teq{L_X\sim 10^{35}} erg/sec, this yields
\teq{\taut\sim 10^{-4}-10^{-3}}, i.e., populations of density \teq{n_e\gtrsim 10^{15}-10^{16}}cm$^{-3}$
that exceed the Goldreich-Julian value \teq{\nGJ = \vert \nabla \cdot \mathbf{E} \vert/4\pi e \sim B/(Pce)} 
by several orders of magnitude \cite{BH07}.
This opacity estimate increases by 2-3 orders of magnitude for scattering at the cyclotron resonance,
which naturally arises in the magnetosphere and defines the high efficiency and the central 
character of the RICS model for magnetar hard tail production.

\begin{figure*}[ht]
\abovecaptionskip=0pt
\centering
\centerline{\hskip -0.2truecm\includegraphics[width=5.7cm]{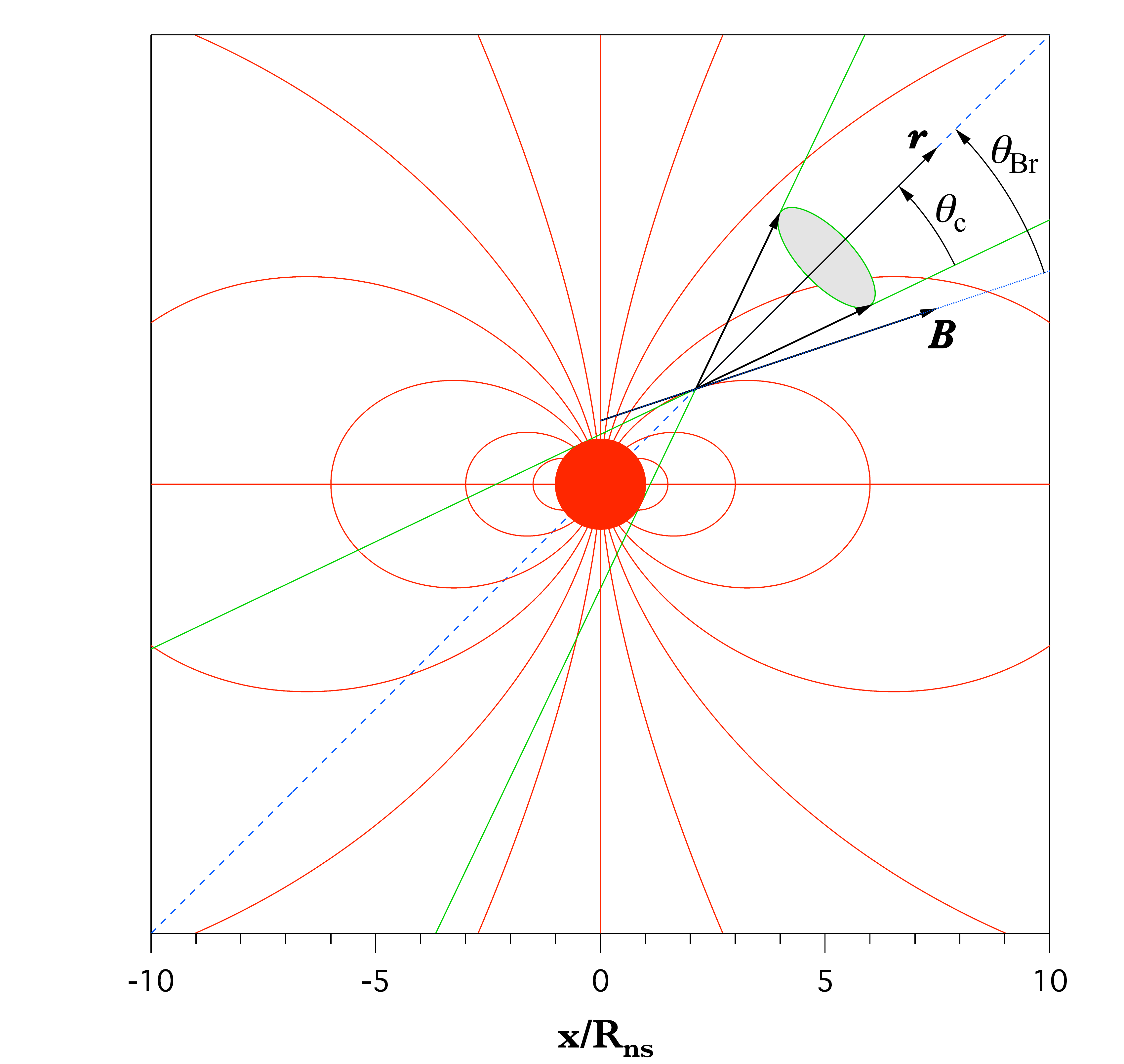}
	\hskip -0.2truecm \includegraphics[width=4.2cm]{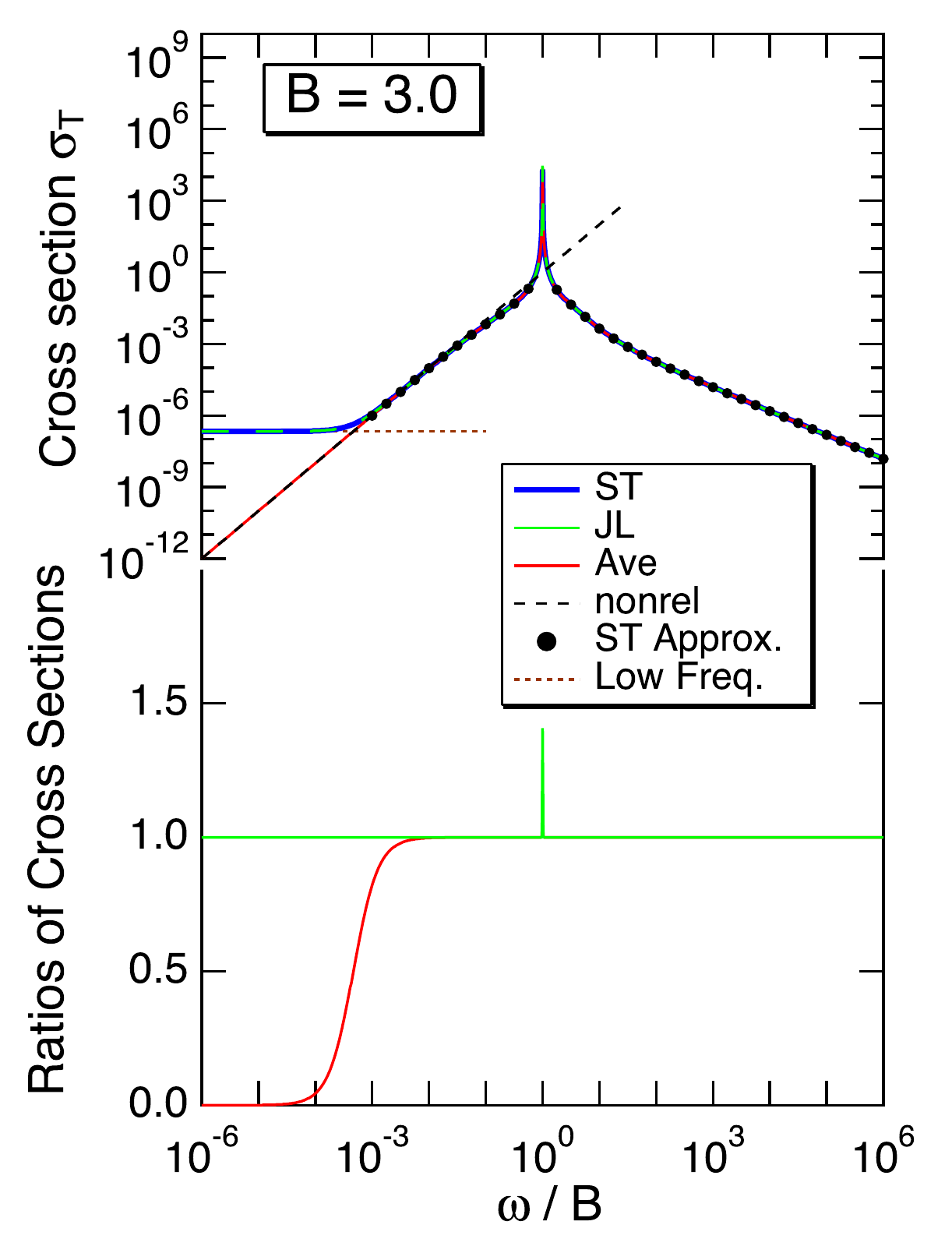}
	\hskip -0.2truecm \includegraphics[width=5.9cm]{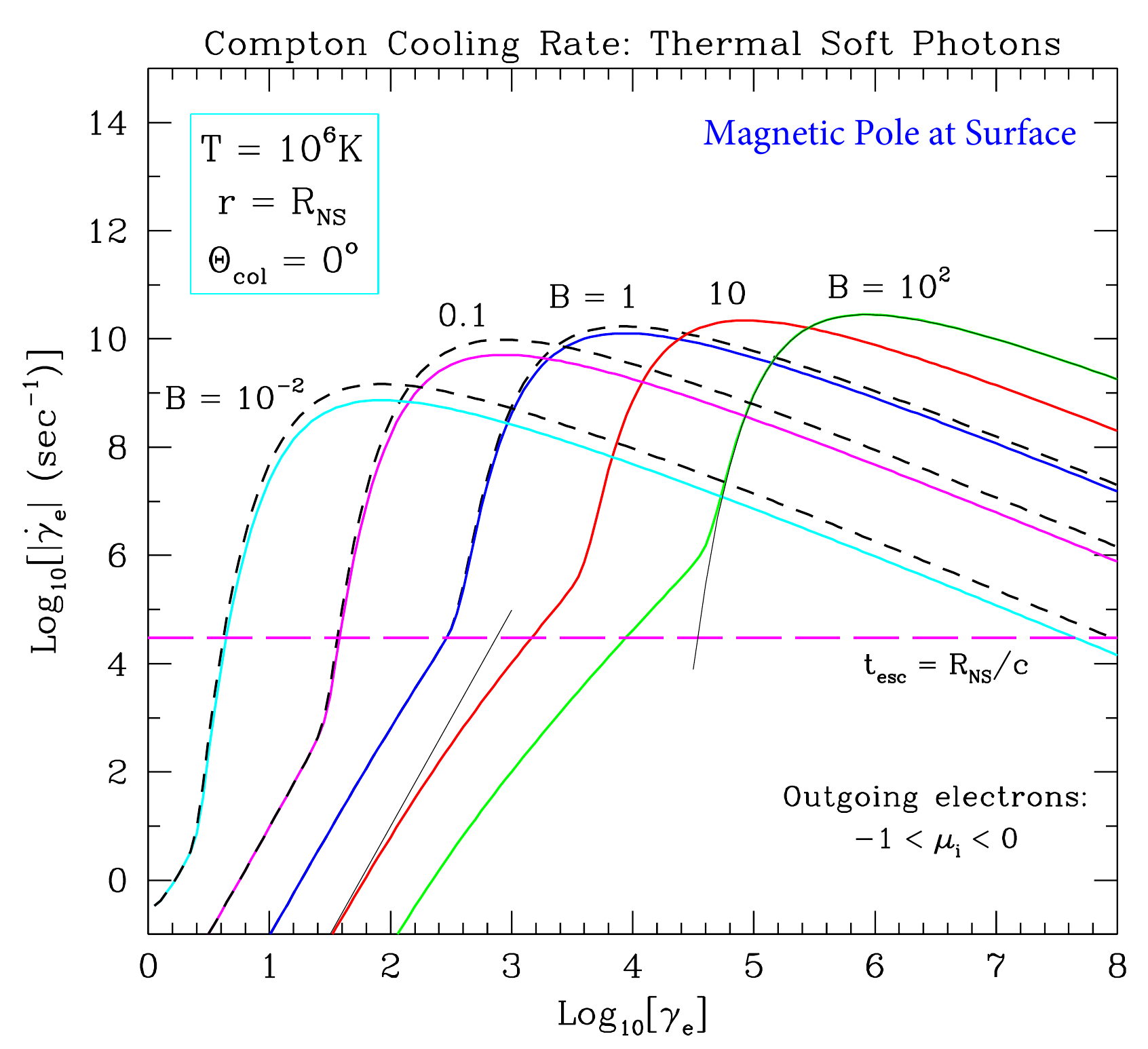}}
\vspace{0pt}
\caption{{\it Left panel}: The magnetospheric geometry for inverse Compton scatterings 
at arbitrary altitudes and colatitudes in the magnetosphere: see \cite{BWG11}.  The green ``cone''
represents the collimated soft photons from the surface for an interaction 
point located at \teq{r/\rns = 3} and a colatitude of \teq{45^{\circ}}. 
The spatial scale is linear, in units of \teq{\rns}.
{\it Center panel}: Total QED magnetic Compton cross sections (in units of \teq{\sigt})
in the ERF  \cite{Gonthier-2014-PRD}, averaged over polarization (upper part), 
in the case of photons moving along {\bf B}.  The spin-dependent 
Johnson \& Lippman (JL; green solid), the Sokolov \& Ternov (ST; blue solid), 
and the spin-averaged (red solid) cross sections are displayed for \teq{B=3} (in units of \teq{B_{\rm cr}}), and in the lower part 
 the ratios of the JL (green) and average (red) cross sections to that of ST 
cross section are displayed.  Only ST is accurate in the cyclotron resonance \teq{\omega \approx B}.
{\it Right panel}: Resonant Compton cooling rates \cite{BWG11} for relativistic electrons 
colliding with thermal X rays (\teq{T=10^6}K) in different fields.  Solid curves are for 
the correct ST choice for electron spinors, dashed are for the JL choice.  The approximate 
Lorentz factor \teq{\gamma_e} of the rapid rise or ``wall'' correlates with 
\teq{B}  because of the approximate resonant scattering condition 
\teq{\gamma_e kT/m_ec^2 \sim B}.  The horizontal dashed line denotes the
light escape scale \teq{1/t_{\rm esc} = c/\rns} corresponding to a stellar radius.
 \label{fig:res_Comp_geom_cool} }      
\end{figure*}

Our group has developed a refined 
upscattering model for this \teq{>10}keV emission over recent years. The nominal geometry for this picture is depicted in 
Fig.~\ref{fig:res_Comp_geom_cool} (left), with relativistic electrons traveling along 
the (red) field lines (\teq{\mathbf{p}_e \parallel \mathbf{B}}) in these slow rotators.  
At a scattering point somewhere on a closed field line in the magnetosphere, 
a relativistic electron energized perhaps by current-driven twisted fields possessing 
toroidal components \cite{TLK02,Beloborodov13,CB17}  
collides with a thermal X-ray (of energy \teq{\erg_sm_ec^2}) 
emanating from the stellar surface within a cone of collimation.
If the kinematic conditions are just right, namely \teq{\gamma_e \erg_s (1 - \cos\thetakB ) = B}, 
the scattering samples a strong resonance at the cyclotron frequency.  
Here \teq{\thetakB} is the angle between the photon momentum and {\bf B}, and hereafter
magnetic field strengths \teq{B} will be expressed in units of 
\teq{B_{\rm cr}\approx 4.41\times 10^{13}}Gauss, the quantum critical
value.   The full QED cross section of Gonthier et al.  \cite{Gonthier-2014-PRD} for \teq{B=3} is depicted 
in the central panel of Fig.~\ref{fig:res_Comp_geom_cool}, and the prominent resonance 
with \teq{\sigma/\sigt \sim 10^2-10^3} is obvious.  The reader can consult \cite{Canuto71PRD,Herold79} 
for the simpler special case of magnetic Thomson scattering.

\noindent{\bf QED Scattering Physics}:
For precision computations of the RICS process in magnetars, accurate full QED formulations 
of magnetic Compton scattering are requisite, including the kinematics of electron recoil and 
Klein-Nishina reductions. State-of-the-art analytics and computations 
of the polarization-dependent, magnetic Compton differential cross section 
\teq{d\sigt/d\Omega} {\it in the electron rest frame} (ERF) have been delivered in \cite{Gonthier-2014-PRD},
and are used in our various modeling papers.  These results incorporate the formally appropriate 
Sokolov \& Ternov (ST) \cite{ST68} spinor formalism, in which the 
wavefunction solutions to the magnetic Dirac equation diagonalize the spin operator \teq{\mu_z} parallel to the field.  
These wavefunctions have become a preferred choice in strong-field QED calculations 
over the last two decades (as opposed to the older Johnson \& Lippmann (JL) \cite{JL49}
eigenstates), since they capture important symmetries.  The ST wavefunctions are symmetric 
between \teq{e^-} and \teq{e^+} states \cite{HRW82,MP83}.   Also, \cite{Graziani93,BGH05} 
established that under Lorentz boosts along {\bf B}, ST states transform
unmixed and yield cyclotron rates that are modified simply via boost Lorentz factors; in contrast,  JL states do not.
The inclusion of a cyclotron decay width \teq{\Gamma} for the intermediate virtual electron state
describes its lifetime, and is essential, rendering the cross 
section finite at the cyclotron resonances.  
This standard Breit-Wigner protocol must employ \underline{spin-dependent} widths 
\cite{HD91,Graziani93,BGH05}, and the 
JL states are not appropriate for correctly implementing such, but 
the Sokolov \& Ternov eigenstates are.  The developments in \cite{Gonthier-2014-PRD}
focused on the singular case of incident photons propagating along {\bf B}, i.e. \teq{\thetakB=0}, 
a suitable ERF specialization for the RICS modeling where \teq{\gamma_e\gg 1} 
introduces relativistic aberration.  In particular, \cite{Gonthier-2014-PRD}
clearly demonstrated the improved precision of the cross section in 
the resonance when employing ST eigenstates (confirmed by \cite{Mushtukov16}), 
revealing the discrepancies incurred when employing 
JL states: for an example, see the middle panel of Fig.~\ref{fig:res_Comp_geom_cool}.

\vspace{3pt}
\noindent{\bf Cooling Rates}:
The focus of our magnetar X-ray tail work first centered on
computations of electron cooling rates due to resonant inverse Compton (IC)
scattering \cite{BWG11} of soft X-rays from entire, isothermal neutron star surfaces.
These rates were derived using ST spinor formalism QED cross sections for scattering,
and clearly demonstrated the inaccuracies inherent in magnetic Thomson evaluations.
The cooling calculations 
were performed in flat spacetime dipole magnetospheres to identify the essential features.   
Sample cooling rates \teq{\dot{\gamma}_e} and their dependence on \teq{\gamma_e} 
and field strength \teq{B_p} at the magnetic polar surface
are illustrated in Fig.~\ref{fig:res_Comp_geom_cool} (right panel).
The forms are essentially inverted images of Planck spectra
governed by the resonant kinematic criterion \teq{\gamma_e\erg_i\sim B}
for seed X rays of energy \teq{\erg_i m_ec^2\sim 3kT\approx 0.25}keV 
(i.e., \teq{\erg_i\sim 5 \times 10^{-4}} for \teq{T=10^6}K).  
The rapid rise in \teq{\dot{\gamma}_e} when \teq{\gamma_e} approaches \teq{B/\erg_i}
can limit putative processes that energize the relativistic electrons in the first place.  
Such radiation reaction would then control the maximum \teq{\gamma_e} of 
electrons, yielding values dependent on the magnetospheric locale.
Fig.~\ref{fig:res_Comp_geom_cool} indicates that generally, resonant
scattering will impose a limit \teq{\gamma_e < 10^5} in polar fields 
\teq{B_p\sim 10^2}.  Moreover, in the lower local fields of \teq{B\lesssim 0.1} 
encountered at higher altitudes and equatorial colatitudes along closed field lines, 
RICS will de-energize fast electrons down to Lorentz factors
\teq{\gamma_e} of the order of 5--20 on lengthscales of \teq{\ell \sim 1-10}cm.
These low \teq{\gamma_e} in turn ascribe low values to the maximum photon energy 
\teq{\sim \gamma_e^2\erg_i} in the IC spectrum, values around \teq{150}keV   
for \teq{\gamma_e\sim 10}.  Accordingly, the observed maximum energies of the 
hard tails in the \teq{<300}keV range might be indicative of the curtailment of 
electron energization by radiation reaction, although it could be caused by 
spectral attenuation: see below.

\begin{figure*}[h]
\abovecaptionskip=0pt
\figureoutpdf{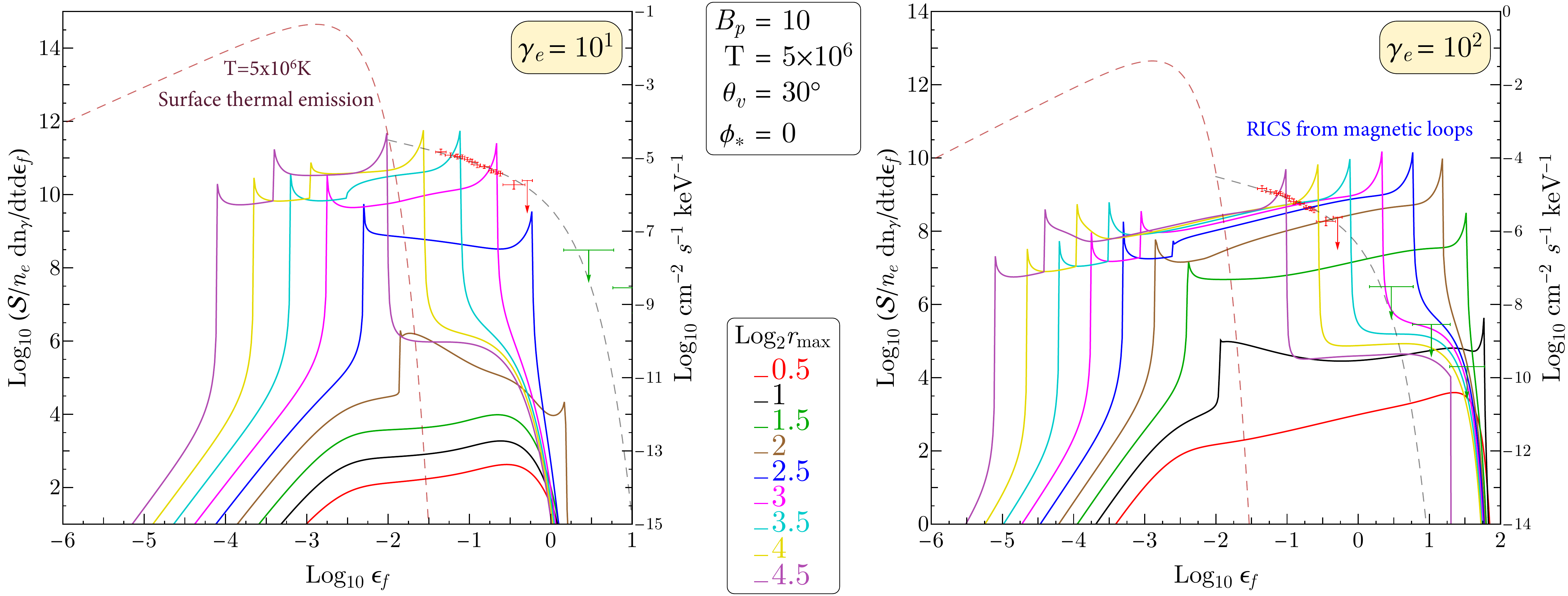}{5.7}{0.0}{0.0}{\rm 
Unnormalized RICS upscattering \teq{F_{\nu}/\nu} spectra for {\it meridional field loops}, 
coplanar with the viewer's direction, depicting nine choices of the maximum loop altitude parameter 
\teq{r_{\rm max} = \{2^{0.5},...,2^{4.5}\}} (in units of \teq{\rns}).  The two panels are for different 
Lorentz factors \teq{\gamma_e}.   
The superposition of these curves gives an indication of spectra that might result from 
toroidal volumes.  Fig.~10 of Wadiasingh et al. \cite{WBGH18}.
Star parameters are the surface polar field \teq{B_p=10} and the uniform surface temperature \teq{T = 5\times 10^6} K.
Spectra are realized for a viewing angle of 
\teq{\theta_v = 30^\circ} to the magnetic axis (a particular pulse phase).
Observational data points \cite{Hartog08b} for AXP 4U 0142+61 are overlaid (red), plus \COMPTEL\ upper limits,
along with a schematic \teq{\erg_f^{-1/2}} power-law with 
a \teq{250} keV exponential cutoff (gray dashed curve).  The Planck spectrum (\teq{kT=0.41}keV)
approximating the soft X-ray data for this magnetar is indicated by the brown dashed curve.
 \label{fig:res_Comp_spec} }      
\end{figure*}

\vspace{3pt}
\noindent{\bf Spectra}:
The next chapter in our studies produced an array of spectral results 
from {\it uncooled} electrons in dipolar field geometry.
Representative spectra for monoenergetic electrons are 
offered in Wadiasingh et al. \cite{WBGH18}, with 
Fig~\ref{fig:res_Comp_spec} showing a \teq{B_p=10} case for two
Lorentz factors \teq{\gamma_e=10, 100}.  The curves therein
(normalized to roughly match the data)
correspond to a fixed viewing angle \teq{\theta_v =30^{\circ}} with respect to 
the instantaneous magnetic dipole axis  \teq{\hat{\muvec}} 
(note that \teq{\theta_v} varies sinusoidally with phase 
as the star rotates).  Therefore these spectra would be sampled 
at a particular rotational phase of the magnetar during its spin period.
This illustration displays what an observer would see from a complete 
closed magnetic field line (footpoint to equator to footpoint) if the line of sight
is coplanar with the field loop (\teq{\phi_0=0^{\circ}}).   This is a rare case
that enables potentially detectable emission out to \teq{\sim\gamma_e^2\erg_i m_ec^2\sim 10}MeV 
when \teq{\gamma_e=100}, for scattering in the Thomson limit (\teq{\gamma_e\erg_i \ll 1}).  
For viewing angles oblique to this plane containing the loop (i.e., \teq{\phi_0 > 0}), 
the spectra are much softer \cite{WBGH18}.
Observe the mismatch between the model spectral slopes
and the data \cite{Hartog08b}.  This is not of concern because the computations were 
for electrons with no energy losses moving along a single loop, and in the absence of 
attenuation processes such as pair creation and photon splitting.
The spectra are parameterized by \teq{\rmax \rns}, the equatorial altitude 
of each magnetic loop, with \teq{\rmax} sampling a logarithmic scale.
The prominent cusps at the highest energies are due to strong Doppler boosting 
of upscattered emission when the line of sight is tangent to a field line,
i.e. parallel to \teq{\mathbf{p}_e}.

The broader ensemble of spectra in \cite{WBGH18}
indicates that AXP/SGR hard tails with cutoffs below 500 keV are best
generated when the observer viewing angles to the magnetic field are
typically greater than \teq{\sim 3^{\circ}}.  
Such a simple conclusion is based in the kinematics of
resonant upscattering. The produced photon energy $\erg_f$ (in units of
$m_ec^2$, and often scaling as $\propto \gamma_e^2 \erg_i$) and the
scattering angle $\theta_f$ with respect to the local field {\bf B} are
strongly correlated via the Doppler boosting condition $\gamma_e\erg_f
(1-\cos\theta_f)\sim B$.  Super-MeV photons are then only observed 
in magnetars when $\theta_f$ is small (typically a few degrees).
Since, for a particular observer perspective that is not along the
magnetic dipolar axis, the local field is seldom tangent to the line of
sight, the emergent spectra in static
magnetospheres then indicate softer emission below around 1 MeV when 
\teq{\gamma_e \lesssim 20}.  These are not too disparate with data 
for the hard X-ray tail cutoffs, as is apparent in the comparison between 
the \teq{\gamma_e=10} and \teq{\gamma_e=100} cases exhibited in 
Fig.~\ref{fig:res_Comp_spec}.  Moreover, the cooling rates depicted at 
the right in Fig.~\ref{fig:res_Comp_geom_cool} suggest that rampant 
resonant Compton cooling at magnetar field strengths may limit electron 
Lorentz factors to around \teq{\gamma_e\sim 10}, indicating approximate 
spectral consistency with the observations.  Note that for viewing
angles approximately above the pole, resonant scattering can proceed at
much higher altitudes where the field is lower (especially if \teq{B_p\lesssim 1}), 
reducing the cooling and increasing the possibly Lorentz factors to 
\teq{\gamma_e > 10^3}.  The associated Doppler boosting should 
precipitate much harder spectra, perhaps consistent with those seen 
in high-field pulsars.

The spectra in Fig.~9 of \cite{WBGH18} are strongly polarized near the maximum 
resonant scattering energies (see also \cite{BH07}) due to the intrinsic
dependence of the Compton cross section on the polarization of the scattered 
photon \cite{Gonthier-2014-PRD}: the \teq{\perp} state always exceeds 
the \teq{\parallel} state.
Here, as always, we adopt a standard linear polarization convention:
\teq{\parallel} (O-mode) refers to the state with the
photon's {\it electric} field vector parallel to the plane containing
{\bf B} and the photon's momentum vector, while \teq{\perp}
(X-mode) denotes the photon's electric field vector being normal to this plane.
This polarimetric signature defines a strong motivation for developing 
future hard X-ray polarimeters.

\begin{figure*}[ht]
\abovecaptionskip=0pt
\centering
\centerline{\hskip -0.2truecm \includegraphics[width=6.0cm]{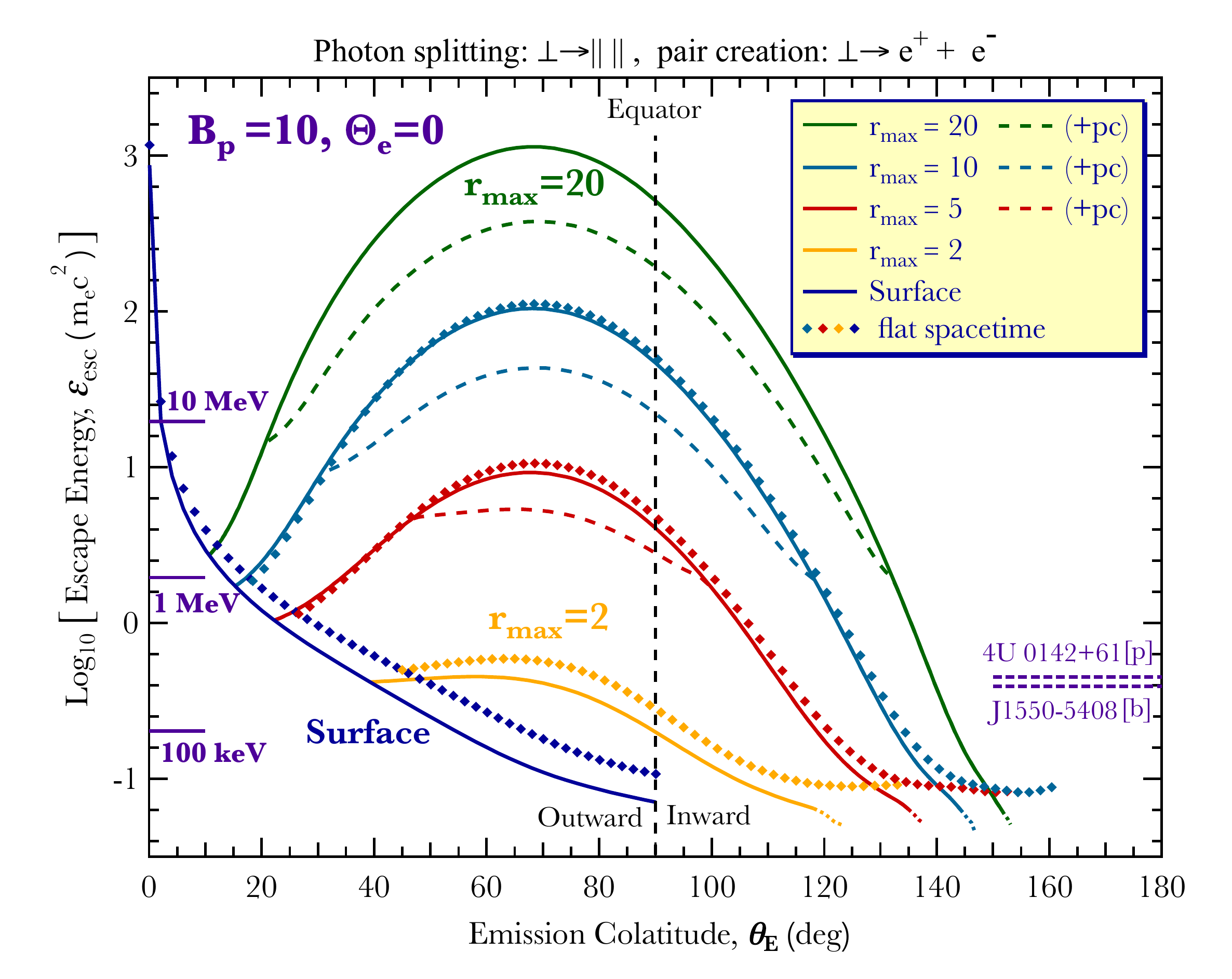}
	\hskip -0.1truecm \raisebox{7pt}{$\includegraphics[width=9.2cm]{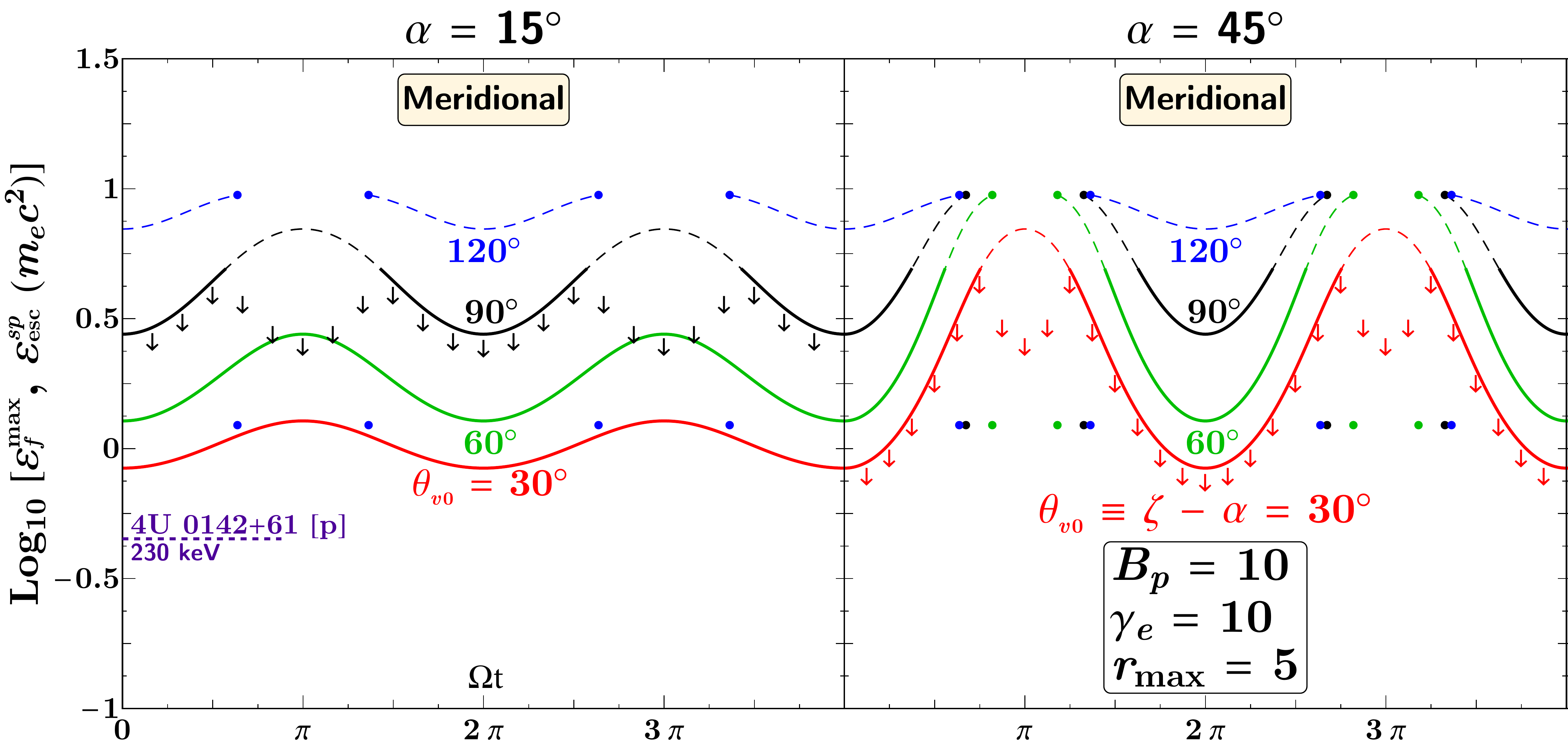}$}}
\vspace{0pt}
\caption{{\it Left panel}: Escape energies for photon splitting (mode \teq{\perp\to \parallel\parallel},
solid curves) and also when adding pair creation of the \teq{\perp} state to the 
total opacity ($+$pc, dashed), for surface polar fields \teq{B_p=10}.
All determinations are for general relativistic propagation, except for those denoted as being for flat spacetime,
namely the traces of diamonds for the surface and \teq{\rmax =2, 5, 10} examples.
A {\it meridional loop} specialization is adopted, with photon emission initially parallel to the local magnetic field line 
(\teq{\Theta_{\rm e} =0}), for maximum loop altitudes  \teq{r_{\rm max}=2,5,10,20} in units of \teq{\rns}.
The surface emission curves are truncated at the equator.
The maximum observed energies for persistent [p] hard tail emission from 4U 0142+61 (230 keV)
and burst emission [b] from SGR J1551-5408 are marked.
{\it Right two panels}: Flat spacetime
photon splitting escape energy \teq{\eescsp} (dotted curves) for
\teq{\perp\to\parallel\parallel} and resonant Compton maximum cutoff energy
\teq{\efmax} (solid and long-dashed) as functions of spin phase \teq{\Omega t},
for oblique rotators with \teq{\alpha \equiv \arccos ( \hat{\Omegavec} \cdot \hat{\muvec} ) = \{15^\circ, 45^\circ \}}.  
Curves are depicted for several choices of the
particular observer viewing angle \teq{\theta_{v0} = \zeta - \alpha} to \teq{\hat{\muvec}} at phase zero
(\teq{\cos \Omega t = 1}), as labeled and color-coded. The \teq{\efmax} emission
energies are depicted as solid curves if \teq{\efmax < \eescsp }, and as long-dashed
loci otherwise, when \teq{\perp} photons are attenuated at
energies below \teq{\efmax}. Curves terminate at dots marking field line footpoint 
emission locales. Downward arrows mark two sample profiles for
effective maximum energies observed in the \teq{\perp} state, signifying
\teq{\min \{ \efmax ,\, \eescsp \}}.  From \cite{Hu19}.}
\vspace{-10pt}
 \label{fig:attenuation}      
\end{figure*}

\noindent{\bf Spectral Attenuation}:
The RICS emission does not necessarily survive to emerge to infinity, 
particularly at energies above 50 keV.  The inner magnetosphere is potentially 
opaque to the escape of such energetic photons due to  
two QED processes that are efficacious in strong magnetic fields: single-photon
pair creation \teq{\gamma\to e^+e^-} and photon splitting \teq{\gamma\to\gamma\gamma}.
The mechanism \teq{\gamma\to e^{\pm}} is active above the \teq{2m_ec^2\approx 1.02}MeV threshold 
\cite{Erber66,DH83,BH01}, is the absorptive part of the birefringent dispersion 
of the magnetized QED vacuum, and is central to the foundation 
of early pulsar models.  Story \& Baring \cite{SB14} computed
$\gamma$-ray opacity for pulsars due to pair creation,
determining that it limits emission to energies below \teq{\sim 10-30}MeV along 
polar field lines when \teq{B_p\gtrsim 10}.  This would explain why the {\it Fermi}-LAT instrument has not detected 
magnetars \cite{Abdo10_Fermi}.  This bound moves to lower energies for equatorial interaction 
zones with increased field line curvature \cite{Hu19}, since the attenuation rate is a strongly 
increasing function of the photon angle \teq{\theta_{\rm kB}} to {\bf B}.
Hu et al. \cite{Hu19} presented a comprehensive 
analysis of pair creation and photon splitting opacity in dipolar magnetar magnetospheres, 
in both flat and curved spacetime. 

Splitting is a higher-order QED 
process, of the order of \teq{\fsc^2 = (e^2/\hbar c)^2} smaller in rate than pair creation \teq{\gamma \to e^{\pm}}.   
It is extremely polarization-dependent so that the interplay 
between scattering and splitting is profound. The splitting modes
\teq{\perp\to\parallel\parallel}, \teq{\parallel\to\perp\parallel} and
\teq{\perp\to\perp\perp} are the only ones permitted by CP invariance in
the limit of zero dispersion \cite{Baring00,BH01}. 
Adler's \cite{Adler71}  selection rules for \teq{\gamma\to\gamma\gamma} argue that 
only the \teq{\perp\to\parallel\parallel} mode of splitting operates in the birefringent, 
magnetized QED vacuum.
Since photon splitting has no threshold imposed by mass creation, it can 
proceed at energies below 1 MeV, and is influential in magnetars down to energies 
of 50 keV or so \cite{BH01}. 
Below pair threshold, the splitting rate scales roughly with energy as \teq{\erg^5} \cite{Adler71,Baring00}.
The trajectory integral analysis of \cite{Hu19}
provided upper bounds of a few MeV or less to the visible energies for magnetars for locales 
proximate to the stellar surface.  This is illustrated in the left panel of Fig.~\ref{fig:attenuation},
which depicts the maximum energy \teq{\eesc} of photons that can escape the magnetosphere 
from select closed field lines and also the stellar surface.  These {\it  escape energies} are depicted
therein as functions of emission colatitude for the \teq{\perp} polarization state only, with 
solid curves for splitting, and dashed curves isolating pair creation when it lowers the escape 
energy at quasi-equatorial colatitudes.  Photons emitted 
in regions within field loops of maximum altitudes \teq{\rmax \sim 2-5} would be attenuated 
by photon splitting at energies below around 250 keV, the highest detected in hard X-ray tails. 
The zones of exclusion are asymmetric between upward/downward emission hemispheres.  

To combine the information on RICS spectroscopy and attenuation escape energies, 
\cite{Hu19} outlined the polarization-dependent effective maximum energies \teq{\efmax} for 
resonant upscattering subject to attenuation by photon splitting (\teq{\eescsp}).  
A subset of the results from Fig.~8 of that work appears in Fig.~\ref{fig:attenuation},
wherein the pulse phase dependence of these energies is depicted.
These were computed for a \teq{B_p=10} magnetar in flat spacetime and a 
magnetic loop with \teq{\rmax=5}.  The two panels 
contrast an almost aligned rotator (\teq{\alpha = 15^{\circ}}) with an oblique one, 
illustrating the increase in modulation with \teq{\alpha}, which is the angle between 
the spin axis \teq{\Omegavec} and the magnetic dipole axis \teq{\muvec}.
The meridional case is for when the electrons traverse magnetic field loops that 
possess locales with tangents pointing toward an observer, 
corresponding to intense and hard radiation signals at particular phases due to strong beaming.
Thus \teq{\efmax} exceeds the observed maxima for the hard tails, and photon splitting 
helps attenuate, but generally only above around 2 MeV.
The contrasting case of anti-meridional motions, where the Doppler beaming
is away from the observer, renders \teq{\efmax} below 100 keV.  Then, photon motions 
are generally inward, and the attenuation by splitting is prolific, suppressing emission 
above 100 keV -- see Fig.~8 of \cite{Hu19}.  The upshot is that sensitive phase-dependent 
spectroscopy+polarimetry above 100 keV will afford tight constraints on the RICS model parameters such as 
\teq{\gamma_e} and the set of field lines that are actively generating the emission.
Such diagnostics may be enabled by the planned medium $\gamma$-ray energy telescope 
AMEGO.\footnote{AMEGO web page: {\it https://asd.gsfc.nasa.gov/amego/index.html}}

\section{Volume-Integrated Spectroscopy and Pulse Profiles}

\begin{figure*}[ht]
\abovecaptionskip=0pt
\centering
\centerline{\hskip -0.2truecm\includegraphics[width=5.7cm]{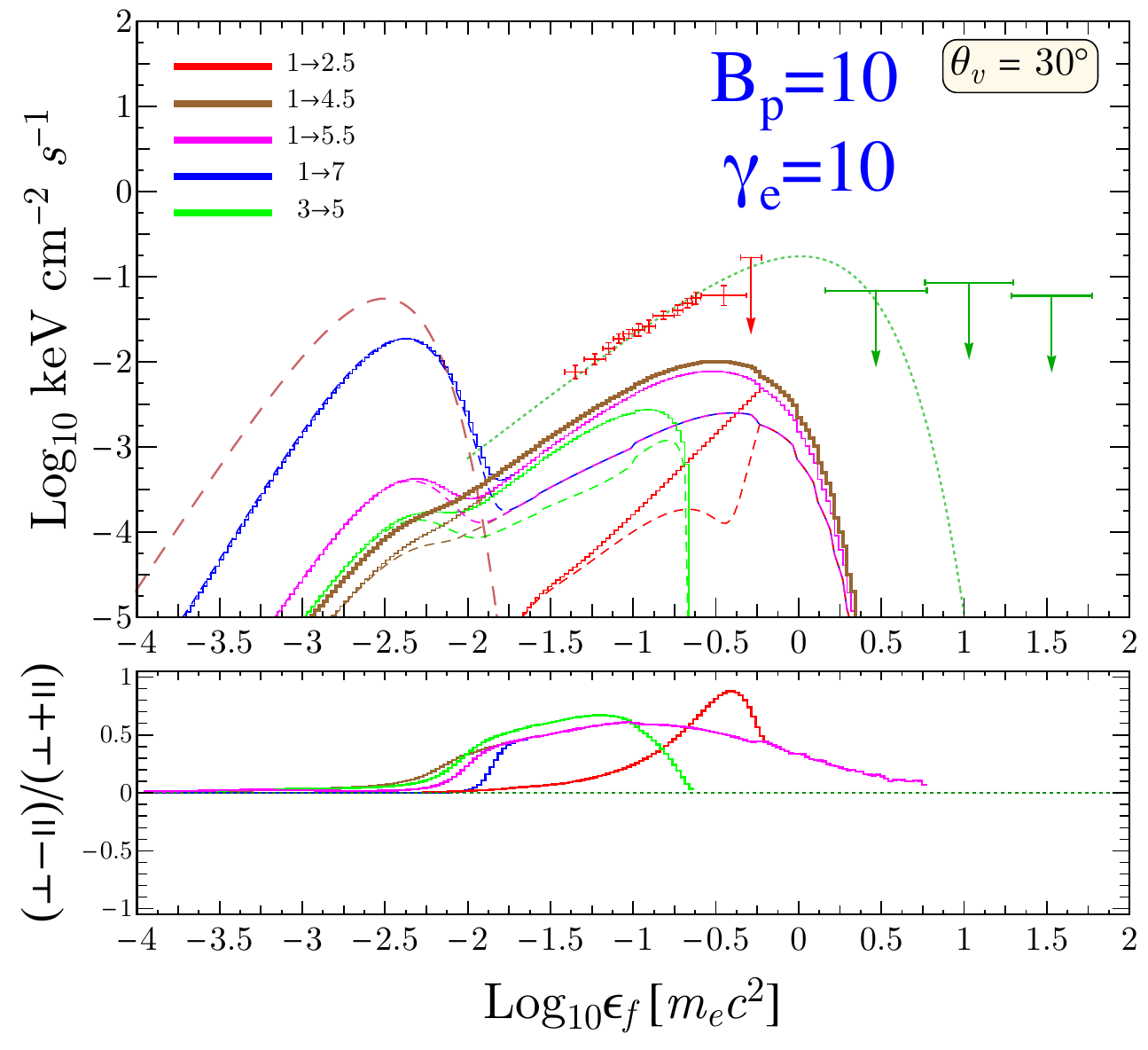}
	\hskip -0.0truecm \includegraphics[width=5.2cm]{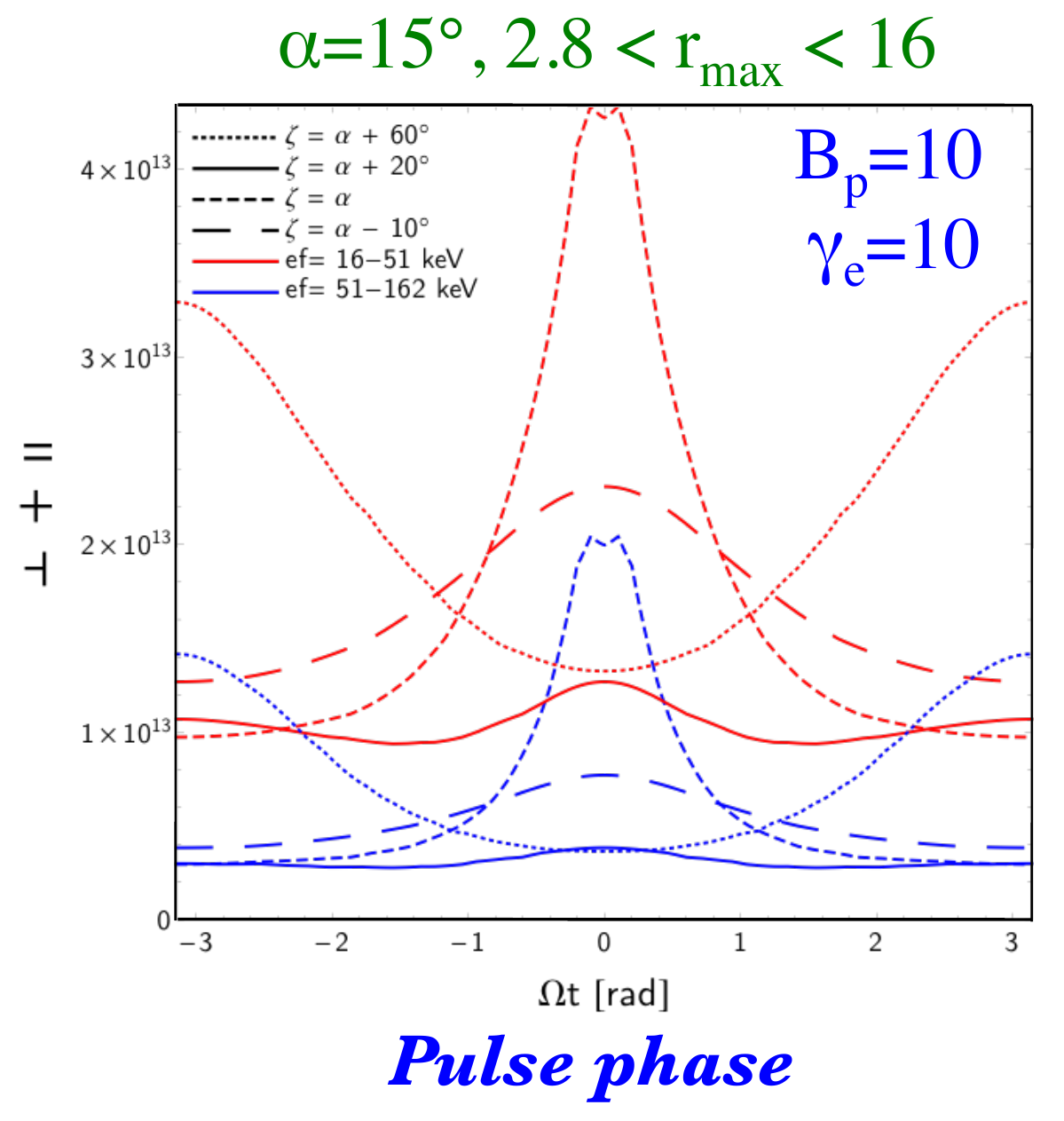}
	\hskip 0.1truecm \includegraphics[width=1.8cm]{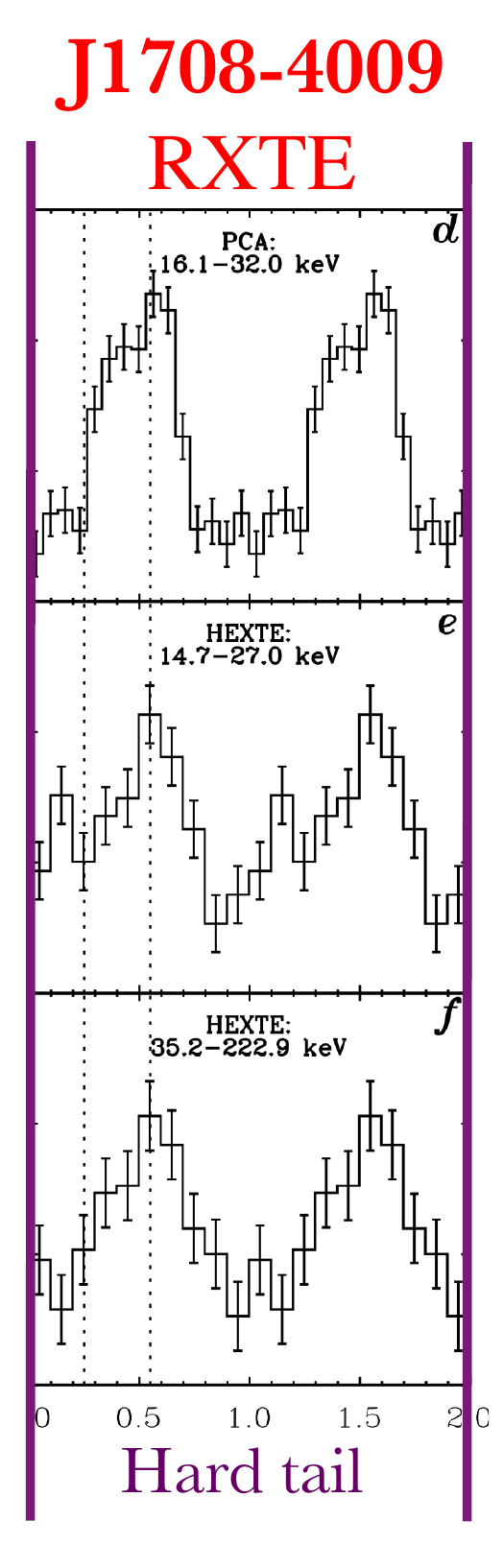}}
\vspace{0pt}
\caption{{\it Left panel}: Polarized inverse Compton spectra (\cite{WBHGHY21}; note the \teq{\nu F_{\nu}} representation) 
for \teq{\gamma_e=10}, from integrations over toroidal volumes of field loops.
These volumes comprise azimuthally-integrated toroidal surfaces specified by \teq{r_{\rm max}=2^{\kappa}}, 
where \teq{\kappa} spans the ranges 
in the legend, each depicted for two polarizations: \teq{\perp} (solid) and \teq{\parallel} (dashed);
the bottom panel displays the polarization degree. 
{\it Center panel}: Pulse profiles for a toroidal volume integrations \teq{r_{\rm max} = 2^{1.5} \to 2^4}, 
generated by a range of \teq{\theta_v} values sampled by a rotating magnetar.  
Red profiles are for \teq{16-51}keV, and 
blue are for \teq{51-162}keV, being obtained for different observer viewing angles
\teq{\zeta} to the rotation axis.  Results are for \teq{B_p=10} and rotator magnetic 
inclination \teq{\alpha = 15^\circ}, and electron Lorentz factor \teq{\gamma_e=10}.
{\it Right panel}: Observed \RXTE\ PCA+HEXTE profiles for different hard X-ray bands 
for RXS J1708-4009 adapted from Fig.~1 of \cite{Kuiper06}.
 \label{fig:spec_pulse_vol_integ} }      
\end{figure*}

The select spectra in Fig.~\ref{fig:res_Comp_spec} from single magnetic field loops 
do not represent an ensemble summation over a bundle of field lines.  The extension 
to integrations of RICS signals over magnetospheric volumes is detailed in a new paper 
Wadiasingh et al. (2021, ApJ in prep. \cite{WBHGHY21}), wherein spectra are obtained using a 
versatile C++ code, again with fixed electron \teq{\gamma_e}, and in flat spacetime.
One of the main results of this new installment of our program is that the 
particular integration over a wide array of field lines determines the spectral index 
below the RICS cutoff or effective maximum energy.  
The principal effect of integrating over toroidal field volumes defined by 
ranges of \teq{\rmax} and magnetic longitudes, as opposed to individual field lines or toroidal surfaces with one 
value of \teq{\rmax}, is to populate lower frequencies more, so that \underline{the spectra steepen}
on average.  An illustrative selection of volume-integrated spectra 
is exhibited in the left panel of Fig.~\ref{fig:spec_pulse_vol_integ}, where results for
five ranges of \teq{\rmax} are depicted.  Attenuation by photon splitting and pair creation 
is omitted from this example, though it is treated at length in \cite{WBHGHY21}.
Corresponding to emission from a hollow toroidal volume, 
these spectra are steeper than those in Fig.~\ref{fig:res_Comp_spec}
(observe the \teq{\nu F_{\nu}} representation).  They therefore match the 4U 0142+61 
\INTEGRAL /\RXTE\ data much more closely when \teq{\gamma_e=10}, particularly 
for \teq{2\leq \rmax \leq 2^{4.5\to 5.5}}, something that doesn't arise when \teq{\gamma_e\gtrsim 100} 
and the \COMPTEL\ upper bounds are violated \cite{WBHGHY21}.  Accordingly, the phase-averaged spectra 
serve as a diagnostic on both the mean Lorentz factor \teq{\gamma_e} {\it and} the 
thickness of the active toroidal volume, at least modulo the assumed dipolar field geometry.
Observe the appearance of prominent low frequency bumps when high altitudes with \teq{\rmax \sim 2^6 - 2^7} are 
incorporated, precipitated by the preponderance of low fields; while obviously interesting, these may be 
precluded by the \teq{<10}keV spectral shape.  Fig.~\ref{fig:spec_pulse_vol_integ} also indicates
a high degree of polarization of the radiation from the scattering, a signature that can be
leveraged by future hard X-ray polarimeters.  

\vspace{4pt}
\noindent {\bf Pulse Profiles}:
A new element of our program is the generation of pulsation profiles for comparison with observations.
Fluxes in any given waveband can routinely be obtained for a rotating star in flat spacetimes.  
The {\it inclination angle} \teq{\alpha} between 
the magnetic dipole \teq{\muvec} and spin axis \teq{\Omegavec} vectors
specifies the rotator geometry.
The observer viewing geometry is prescribed via the {\it aspect angle} 
\teq{\zeta} between the line of sight {\bf O} and spin axis \teq{\Omegavec} vectors.
Angles \teq{\alpha} and \teq{\zeta} are standard stellar parameters for pulsar studies.
The instantaneous viewing angle \teq{\theta_v} varies in a sinusoidal fashion as the 
magnetar rotates \cite{WBGH18}, with its average corresponding to \teq{\alpha}, and its amplitude 
being \teq{\vert \zeta - \alpha \vert}.  Spectra analogous to those in Fig.~\ref{fig:res_Comp_spec}
are obtained for an array of \teq{\theta_v}, integrated over set frequency ranges and 
presented as intensity ``sky maps'' in \cite{WBHGHY21}; these maps have 
pulse phase \teq{\Omega t} on the abscissa, and \teq{\zeta} as the ordinate.  Horizontal cuts through 
these sky maps produce pulse profiles, and we exhibit such in the center panel of 
Fig.~\ref{fig:spec_pulse_vol_integ} for a rotator inclination \teq{\alpha=15^{\circ}}.  
These profiles, which in general depend on the thickness of the toroidal emission volume, 
clearly display first and second-order Fourier 
components that result from the azimuthally-symmetric and hemispherically-symmetric 
volumes adopted in the integration.  Real pulse profiles such as those for RXS J1708-4009 
on the right of Fig.~\ref{fig:spec_pulse_vol_integ} display more harmonic structure, 
providing evidence for departures from these spatially-congruent specializations.  A future stage of our 
program will consist of an exploration of these subtleties, using the pulse profile data 
to inform and constrain the azimuthal/longitudinal dimension of the radiating toroidal volume.
Yet, with just the information presented
in Fig.~\ref{fig:spec_pulse_vol_integ}, it is evident 
that pulse profile comparison between model and data provides powerful diagnostics 
on stellar parameters \teq{\alpha} and \teq{\zeta}, as well as the \teq{\rmax} range.
Notably, pulse profiles 
with simple quasi-sinusoidal traces generally preclude values of \teq{\alpha \gtrsim 40^{\circ}}
as they generate higher Fourier components \cite{WBHGHY21}
due to visibility of emission from both hemispheres.

\section{Conclusion}

The results surveyed here from our extensive program exploring resonant inverse Compton 
scattering models for magnetar hard X-ray tails display the evolution and depth of the analyses, 
the complexity of the modeling.  The considerable sophistication of these undertakings 
will be further enhanced when the spectral and temporal emission information is combined with concurrent
RICS cooling of the electrons.  This serves as the next stage of this enterprise, which is 
well-positioned to afford observational diagnostics with extant pulse profile and spectral data 
from \RXTE, \INTEGRAL, \NuSTAR\ and {\it Fermi}-GBM.  Yet the polarimetric element of 
our studies is an attractive antecedent to a future era of hard X-ray polarimetry, when even more 
powerful probes of the mysteries of magnetar magnetospheres will be enabled.
Foremost among these is the prospect of being able to experimentally demonstrate the 
verity of magnetic photon splitting and single-photon pair creation, and the 
intimately-connected birefringence of the magnetized quantum vacuum, heretofore 
untested theoretical predictions for high-field QED domains.  
Magnetars thus serve as a potent cosmic QED physics laboratory, and the model 
developments will foster this advance by disentangling source emission and geometry
information from the signatures of strong-field QED physics.

\vspace{5pt}
\acknowledgements{M.~G.~B. acknowledges the generous support of the 
National Science Foundation through grant AST-1517550, and 
NASA's {\it Fermi} Guest Investigator Program through grant NNX16AR66G. 
Z.~W. is supported by the NASA postdoctoral fellowship program.}
\vspace{-5pt}

\renewcommand{\baselinestretch}{0.92}

\end{document}